\documentclass[journal]{IEEEtran}

\usepackage{graphicx}
\usepackage{multirow}
\usepackage{multicol}
\usepackage{threeparttablex}
\usepackage{pifont}
\usepackage{mathtools}
\usepackage[flushmargin]{footmisc}
\usepackage{cite}
\usepackage{subfig}

\newcommand{\cmarkn}{\ding{51}}
\newcommand{\xmark}{\ding{55}}

\title{Identifying malicious accounts in Blockchains using Domain Names and associated temporal properties}
\author{
\IEEEauthorblockN{Rohit Kumar Sachan, Rachit Agarwal, Sandeep Kumar Shukla}\\
\IEEEauthorblockA{
 CSE Department, IIT Kanpur\\
 Email: \{rohitks, rachitag, sandeeps\}@iitk.ac.in}
}
\begin{document}

\maketitle

\begin{abstract}

The rise in the adoption of blockchain technology has led to increased illegal activities by cyber-criminals costing billions of dollars. Many machine learning algorithms are applied to detect such illegal behavior. These algorithms are often trained on the transaction behavior and, in some cases, trained on the vulnerabilities that exist in the system. In our approach, we study the feasibility of using metadata such as Domain Name (DN) associated with the account in the blockchain and identify whether an account should be tagged malicious or not. Here, we leverage the temporal aspects attached to the DNs. Our results identify 144930 DNs that show malicious behavior, and out of these, 54114 DNs show persistent malicious behavior over time. Nonetheless, none of these identified malicious DNs were reported in new officially tagged malicious blockchain DNs. 
\end{abstract}

\vspace{-0.3cm}
\section{Introduction}

Cyber-criminals/attackers always explore new ways to perform malicious activities over the Internet. Malicious activities are either socially engineered (like Phishing, Spam) or exploit vulnerabilities in a system (like hacks) with a motivation to either steal information, sabotage operations, or corrupt the hardware/software to earn illegitimate revenue such as cryptocurrency. Cryptocurrencies are based on blockchain technology, where cyber-criminals leverage the anonymity aspects of the blockchain technology for the malicious activities~\cite{cbronline}.

In the infrastructure supporting blockchain, especially permissionless blockchains such as Ethereum, many organizations that are involved in cyber-crime, create malicious distributed applications (dApps) and malicious blockchain wallets (BCWs) that people sometimes use (deliberately or accidentally) and get victimized. One such malicious activity is that of \textit{Gambling} which is illegal in many countries such as India. In Ethereum, dApps are programs (commonly called Smart Contract (SC) that are written in `Turing complete' Solidity language), and BCWs are digital wallets (typically used to store and manage cryptocurrency). Both dApps and BCWs have associated addresses. The dApps and BCWs interact with other SCs, wallets, and other externally owned accounts (EOAs - accounts owned by people or entities) to perform illegal transactions~\cite{hackernoon}. Thus, it is highly desirable to efficiently and effectively detect and report such dApps and BCWs and safeguard the blockchain users. 

Most of the related state-of-the-art malicious account detection techniques include the transaction-based features~\cite{agarwal2020detecting, agarwal2021detecting} or study vulnerabilities present in the blockchain ecosystem~\cite{maleh2020blockchain}. Malicious dApps and BCWs, besides their transaction behavior, have unique signatures where their metadata shows different characteristics. Such metadata includes Domain Name (DN) and associated Domain Name System (DNS) records (including IP address, IP prefix, and NS address). DNs are human-readable alias names given to Internet-based services running on IP addresses~\cite{blockgeeks}. Thus, we ask, \textbf{\textit{can such DN-related information be helpful in the identification of malicious dApps and BCWs?}} There are two ways to answer the above question \textit{(i)} we include DN related features in the current ML-based approaches and study the improvements, and \textit{(ii)} as DNs are associated with accounts, we only use DN related features, use ML to identify malicious DN, and assume that the associated account is also malicious. In this work, we only focus on the DNs and not on the transactions of a particular dApp or a BCW to identify malicious dApps and BCWs. Note that Ethereum provides an Ethereum Name Service (ENS) where users, including dApps and BCWs can associate DNs to their addresses. The ENS domains are different from traditional DNs as they do not have associated DNS records. In Ethereum, Zedrun\footnote{Zedrun: https://zed.run/, Note that we have no intention to align any transaction of with Zedrun, it is just an example for us.} is an example of a gambling dApp which does not have an ENS entry on Ethereum, but it has DN and DNS records.

Nonetheless, in many cases, a malicious DN is algorithmically generated and shows patterns. There are various state-of-the-art techniques to detect malicious DNs. In~\cite{sahoo2017malicious}, the authors present a survey of state-of-the-art approaches (including blacklist-based and Machine Learning (ML) based approaches) that detect malicious DNs. However, these methods have drawbacks. Blacklist-based methods cannot adapt to the behavioral changes and capture any hidden aspects. While ML-based techniques use various features, but in most cases do not provide any information on whether their hyperparameters are optimized. Nonetheless, it requires continuous monitoring of DNS traffic (DNs request-responses and associated DNS records) from the behavior and social perspective to detect malicious DN in a blockchain ecosystem. The state-of-the-art algorithms do not use such temporal aspects.
 
This motivates us to include temporal features and non-temporal features in the ML pipeline to detect malicious DNs and tag associated dApps and BCWs as \textit{suspects}. We call the identified malicious dApps and BCWs as \textit{suspects} because these accounts might not be officially tagged, but their features might show behavioral similarity with the already marked malicious accounts. Here, we first identify the temporal properties that exist by analyzing the temporal behavior of the DNS traffic data. On top of the features that are lexical-based, DNS query-based, host-based, and static graph-based, we introduce the DNS traffic-based temporal features. These features are burst-based (such as inter-event time burst and query frequency burst) and temporal graph-based (such as changes in the degree and diameter of the underlying network formed using IP address, NS address, and IP prefix associated with the DNS records). Thus, including both temporal and non-temporal features in our study.  Using such a rich feature set, we only study the behavior of DNs over two different temporal (2 hr based and 1 month based) granularities. The reason for such a choice is the computational resources available to us. Using the feature set, we study both supervised and unsupervised ML algorithms for completeness. Our approach using supervised learning is able to detect malicious blockchain DNs with 89.53\% balanced-accuracy and all malicious DNs (including malicious blockchain DNs) with 81.52\% balanced-accuracy. 
While when using unsupervised learning, our approach detects 144930 DNs that show malicious behavior at least once in a given temporal granularity and 54114 DNs that show persistent malicious behavior but are not marked. 

With respect to the blockchain technology, to the best of our knowledge, ours is the first work that identifies the significance of temporal aspects of DNs and analyzes them to detect suspect BCWs and dApps in a blockchain. In summary, our core contributions through this work are:

\begin{itemize}
  \item \textbf{Comparative study}: We present a comparative study of the various state-of-the-art techniques used towards detecting the malicious DNs. Here we present the features the use, what they detect, ML algorithms they use, what dataset they use, issues with their approach, hyperparameters of the used ML algorithms, and reported performance. 
  
  \item \textbf{Features:} We identify the feature vector that includes temporal and non-temporal features. The temporal features are based on the time-series analysis of the DNS traffic data and are based on burst, degree, and diameter. 
  
  \item \textbf{Behavior analysis:} We analyze the temporal behavior of the malicious DNs, benign DNs, and malicious DNs associated with the blockchain and apply ML-based approach to detect malicious DNs. We find that DecisionTree performs the best in terms of balanced-accuracy for our dataset. Our approach, based on Unsupervised ML algorithm, reports 54115 DNs that are not marked as malicious, but their behavior is similar to the malicious DNs. These DNs were also not present in the new list of malicious DNs that were identified between 30$^{st}$ October 2020 and 7$^{th}$ April 2021. Nonetheless, we cannot identify which DNs are associated with accounts present in a blockchain due to security reasons.


\end{itemize}

The rest of the paper is organized as follows. In Section~\ref{sec:RW}, we present the state-of-the-art works related to the detection of malicious DNs using DNS traffic data. In Section~\ref{sec:method}, we present our methodology and followed by in-depth evaluation along with the results in Section~\ref{sec:results}. We finally conclude in Section~\ref{sec:conc}.

\vspace{-0.3cm}
\section{Related Work}\label{sec:RW}

This section discusses the state-of-the-art approaches that target the identification of malicious DN using ML algorithms. Based on our survey on different state-of-the-art techniques to detect malicious DNs, we identify that no approach targets blockchain accounts and uses temporal features associated with DNs of the dApps and BCWs. They talk about DNs in general. Generally, the state-of-the-art techniques mainly use features extracted after studying either DN strings, underlying static graph, or DN query.


Analysis of substrings present in a DN is one of the most popular way to classify it as malicious. In~\cite{hoang2018botnet}, the authors present an ML-based model to detect DNs associated with botnets. The model uses statistical characteristics extracted using N-grams and the vowel distribution characteristics in a DN. N-gram is a Natural Language Processing (NLP) concept where a string is segmented into substrings up to length N. Then, the weights of each substring is calculated based on the frequency of occurrence of each substring. This approach has a drawback. In N-grams, the number of features increases exponentially with respect to ‘N’. To overcome such an issue, in~\cite{selvi2019detection}, the authors propose a masked N-Gram approach where symbols substitute each character, i.e., character `c' replaces all consonant, `v' substitute all vowels, `n' substitutes all digits and `s' substitutes all others characters in the given string. Such substitution reduces the number of features significantly. In many cases, attackers use algorithmically generated DNs to avoid detection by such N-gram-based techniques. In~\cite{anand2020ensemble}, the authors present an ML-based approach to detect such malicious DNs that are algorithmically generated (mainly using a Domain Generation Algorithm (DGA)). The approach uses lexical and statistical features (such as N-gram and character occurrence frequency in DNs). In follow-up work, in~\cite{sai2020DGA}, the authors present an ML-based method for detecting the malicious Word-list-based DGA DNs based on lexical and network-based features (such as created since, updated since, registrar, and Time-to-Live  (TTL)). 


Besides algorithmically generated DNs, attackers use the concept of changing IPs (i.e., IP-Flux or Fast-Flux) to perform malicious activities such as phishing~\cite{zhauniarovich2018survey}. Such DNs are hard to detect. In~\cite{perdisci2009detecting,FluxBusterperdisci2012early}, the authors present an ML-based approach to detect malicious Fast-Flux DNs that use features (like short TTL and the high number of resolved IPs) available from the passive DNS traffic data. In another approach, in~\cite{stalmans2011framework}, the authors detect malicious DNs that are DGA generated and have Fast-Flux behavior after analyzing the DN strings and the associated DNS query records. In a similar work, in~\cite{truong2020detecting}, authors present an ML-based method to classify a Fast-Flux-based DN as malicious or benign. Their approach uses an extended feature set (i.e., uses features such as the number of resolved IP address, the minimum value of TTL, entropy of prefixes, entropy of sub-domains, stability of IP address pool, the time between first seen and last seen) to improve the confidence interval to 95\%. In~\cite{messabi2018malware}, the authors present an URL-based malicious DNs detection system. Their system uses DN-based features and DNS query-based features. They study past DNS activities of each DN and explore the difference between the physical behavior of benign and malicious DNs.  


Many approaches also assign a reputation score to classify a DN as malicious or benign. In~\cite{zhao2019malicious}, the authors use an N-gram concept to calculate the reputation value of a DN. They calculate the reputation value of a DN based on its weight value in the DN’s substrings. The algorithm marks all the DNs whose reputation value is above a threshold as malicious DN. Similarly, in~\cite{Notosantonakakis2010notos}, the authors present \emph{Notos}, a dynamic reputation system for DNs. Notos assigns a low reputation score to a malicious DN based on the characteristics of the malicious DN that include resource provisioning, usages, and DN management. In~\cite{Exposurebilge2011exposure, Exposurebilge2014exposure}, the authors present \emph{Exposure}, which assigns a reputation score based on the analysis of passive DNS data for detecting the malicious DNs. Exposure identifies behavioral differences between benign and malicious DNS based on time-based features (lifetime, repeating patterns), DNS answer-based features (unique IP addresses and countries, number of DNs associated with an IP), TTL value-based features (unique TTL, number of TTL changes), and DN-based features (string-based) that exists in the DNS traffic data.
Both Notos and Exposure need initial IP reputation. However, in cases when such reputation is not present, both Notos and Exposure fail to perform. To overcome this issue, in~\cite{Kopisantonakakis2011detecting}, the authors present \emph{Kopis} that monitors the DNS traffic between the top-level domains (TLD) and authoritative name servers (AuthNSs). 


Other malicious DN detection systems use graph data structure (i.e., underlying associated DNS graph). These systems use graph properties, node connectivity, and edge association to identify malicious DNs. In~\cite{stevanovic2015ground}, the authors propose an ML-based approach to obtain labels of the DNs based on the features selected from graph components by monitoring the malicious DNS traffic. In another work~\cite{khalil2016discovering}, the authors present a non-ML malicious DNs detection approach using the DN-resolution graph. The approach assumes that if a DN is strongly associated (connected) with a known malicious DN, it will likely be malicious. In~\cite{Segugiorahbarinia2016efficient}, the authors present \emph{Segugio}, a malware-control DN system. It closely monitors the DNS query behavior of malware-infected machines and builds a machine-domain bipartite graph that represents ‘who is querying what'. It detects the previously unknown malware DNs based on the query behavior of DNs. In~\cite{tran2017dns}, the authors construct a DNS graph between DN and IP address for only A-records and use connectivity between DNs and malicious DNs to tag DNs as malicious. Their performance purely depends on the association/link with known malicious DNs.
    

Note that all the above ML-based techniques have inherent data imbalance problems. The number of malicious DNs is low as compared to benign DNs. In~\cite{liu2018imbalanced}, the authors present the \emph{HAC\_EasyEnsemble}-based model to overcome the data imbalance problem. This model extracts static lexical features and dynamic DNS resolving features to profile every DN from the DNS traffic data. In~\cite{wang2020malicious}, the authors also address the imbalance problem and present a \emph{KMSMOTE} method that uses SMOTE 
and K-Means clustering algorithm. The system uses assumptions such as malicious DNs leave their traces on DNS traffic, malicious DNs have lower DN registration cost, and reuse network resources. 


In general, the ML-based approach to detect malicious DNs have high time complexity. In~\cite{shi2018malicious}, the authors present a method to detect malware DNs using Extreme Learning Machine (ELM) to improve time complexity. ELM is new, fast learning, and highly accurate Neural Network (NN) scheme for Single-hidden-Layer Feed-forward NNs (SLFNs).

We summarize these state-of-the-art techniques in more detail in Table~\ref{table:rw}. Here, we report the features that these approaches use, what they detect, which ML algorithms they test, what dataset they use, and the issues these approaches have. In Table~\ref{table:hy}, we report the hyperparameters of the ML algorithms that these approaches have used along with the reported performance. Most of the state-of-the-art approaches use the supervised learning method and do not report the hyperparameters they use to train the learning algorithms. We also observe that all the state-of-the-art approaches use features that do not capture the temporal behavior of DNS data. To the best of our knowledge, none of the approaches uses temporal (or time series) features and do not target identifying malicious dApps and malicious BCWs using DNs.

\begin{table*}
\vspace{-0.5cm}
\centering
\caption{Features used in related studies}
\label{table:rw}
\begin{threeparttable}

\begin{small}
\begin{tabular}{|c|c|ccccc|c|c|c|c|c|c|}
    
    \hline
    \multirow{2}{*}{\#} & \multirow{2}{*}{B/C}  & \multicolumn{5}{c|}{Features Used} & \multicolumn{2}{c|}{Detection}  & \multirow{2}{*}{ML Algo.} & \multicolumn{2}{c|}{Datasets} & \multirow{2}{*}{Issues}\\
    \cline{3-9}\cline{11-12}
    
    & & N & Q & G & T & O & Of & Tackles &  & Sources used & Size &   \\
    \hline
    
    \multirow{2}{*}{\cite{hoang2018botnet}} & 
    \multirow{2}{*}{\xmark} & 
    \multirow{2}{*}{\cmarkn} & 
    \multirow{2}{*}{\xmark} & 
    \multirow{2}{*}{\xmark} & 
    \multirow{2}{*}{\xmark} & 
    \multirow{2}{*}{\xmark} &
    \multirow{2}{*}{BT} &
    \multirow{2}{*}{-} & 
    KNN, C4.5, & 
    \multirow{2}{*}{\cite{alexa1m,dgadata,conficker}} & 
    \multirow{2}{*}{60K} & \\
    & & & & & & & & & RF, NB & & & \multirow{1}{*}{Only based}\\
    \cline{1-12}

    \multirow{1}{*}{\cite{selvi2019detection}} & 
    \multirow{1}{*}{\xmark} & 
    \multirow{1}{*}{\cmarkn} & 
    \multirow{1}{*}{\xmark} & 
    \multirow{1}{*}{\xmark} & 
    \multirow{1}{*}{\xmark} & 
    \multirow{1}{*}{\xmark} & 
    \multirow{1}{*}{AG} & 
    \multirow{1}{*}{-} & 
    \multirow{1}{*}{RF} & 
    \multirow{1}{*}{\cite{alexa1m} \cite{dga}} & 
    \multirow{1}{*}{64K} & 
    \multirow{1}{*}{on}\\
    \cline{1-12}
    
    \multirow{3}{*}{\cite{anand2020ensemble}} & 
    \multirow{3}{*}{\xmark} & 
    \multirow{3}{*}{\cmarkn} & 
    \multirow{3}{*}{\xmark} & 
    \multirow{3}{*}{\xmark} & 
    \multirow{3}{*}{\xmark} & 
    \multirow{3}{*}{\xmark} & 
    \multirow{3}{*}{AG} & 
    \multirow{3}{*}{-} & 
    C5.0, GB,& 
    \multirow{3}{*}{\cite{alexa1m,dgadata}} & 
    \multirow{3}{*}{51K} & string analysis\\
    & & & & & & & & & RF, CART,& & &\\
    & & & & & & & & & KNN, SVM & & & \\
    \cline{1-13}

    \multirow{3}{*}{\cite{sai2020DGA}} & 
    \multirow{3}{*}{\xmark} & 
    \multirow{3}{*}{\cmarkn} & 
    \multirow{3}{*}{\xmark} & 
    \multirow{3}{*}{\xmark} & 
    \multirow{3}{*}{\xmark} & 
    \multirow{3}{*}{\cmarkn} & 
    \multirow{3}{*}{AG} & 
    \multirow{3}{*}{-} & KNN, GB, RF, & 
    \multirow{3}{*}{\cite{alexa1m,dgadata,whois}} & 
    \multirow{3}{*}{40K} & 
    \multirow{1}{*}{Based on limited}\\
    & & & & & & & & &  CART, C4.5, & & & DGA based \\
    & & & & & & & & &  NB, C5.0 & & & DNs \\
    \hline

    \multirow{2}{*}{\cite{perdisci2009detecting}} & 
    \multirow{2}{*}{\xmark} & 
    \multirow{2}{*}{\xmark} & 
    \multirow{2}{*}{\cmarkn} & 
    \multirow{2}{*}{\xmark} & 
    \multirow{2}{*}{\xmark} & 
    \multirow{2}{*}{\cmarkn} & 
    \multirow{2}{*}{FF} & 
    \multirow{2}{*}{-} & 
    \multirow{2}{*}{C4.5} & 
    \multirow{2}{*}{American ISP} & 
    \multirow{2}{*}{2.5B} & \multirow{1}{*}{Limited to fast-flux,}\\
    & & & & & & & & & & & & \multirow{1}{*}{IP and TTL based}\\
    \hline
    
    \multirow{2}{*}{\cite{FluxBusterperdisci2012early}} & 
    \multirow{2}{*}{\xmark} & 
    \multirow{2}{*}{\xmark} & 
    \multirow{2}{*}{\cmarkn} & 
    \multirow{2}{*}{\xmark} & 
    \multirow{2}{*}{\xmark} & 
    \multirow{2}{*}{\cmarkn} & 
    \multirow{2}{*}{FF} & 
    \multirow{2}{*}{-} & 
    \multirow{2}{*}{C4.5} & 
    Security Info. & 
    \multirow{2}{*}{\xmark} & 
    \multirow{1}{*}{Based on local} \\
    & & & & & & & & & & Exchange & & RDNS traffic \\
    \hline

    \multirow{2}{*}{\cite{stalmans2011framework}} & 
    \multirow{2}{*}{\xmark} & 
    \multirow{2}{*}{{\cmarkn}} & 
    \multirow{2}{*}{{\cmarkn}} & 
    \multirow{2}{*}{\xmark} & 
    \multirow{2}{*}{\xmark} & 
    \multirow{2}{*}{\cmarkn} & 
    FF, & 
    \multirow{2}{*}{-} & 
    \multirow{2}{*}{C5.0, BC, NB}  & 
    \cite{fastfluxtracker,googledoubleclick}, & 
    \multirow{2}{*}{40M} & 
    \multirow{1}{*}{No mention of} \\
    & & & & & & & AG & & & \cite{spyeye,zeustracker} & & \multirow{1}{*}{evolution of DNs} \\
    \hline

    \multirow{3}{*}{\cite{truong2020detecting}} & 
    \multirow{3}{*}{\xmark} & 
    \multirow{3}{*}{\xmark} & 
    \multirow{3}{*}{\cmarkn} & 
    \multirow{3}{*}{\xmark} & 
    \multirow{3}{*}{\xmark} & 
    \multirow{3}{*}{\cmarkn} & 
    \multirow{3}{*}{FF} & 
    \multirow{3}{*}{-} & 
    \multirow{3}{*}{RF} & \cite{alexa1m,cybercrimetracker,hpHosts},  & 
    \multirow{3}{*}{5.4M} & 
    \multirow{1}{*}{Only based on} \\
    & & & & & & & & & & \cite{malwaredomainlist,phishtank}, & & n/w and IP \\
     & & & & & & & & & & \cite{zeustracker,spyeye} & & based features \\
    \hline

    \multirow{2}{*}{\cite{messabi2018malware}} & 
    \multirow{2}{*}{\xmark} & 
    \multirow{2}{*}{\cmarkn} & 
    \multirow{2}{*}{\cmarkn} & 
    \multirow{2}{*}{\xmark} & 
    \multirow{2}{*}{\xmark} & 
    \multirow{2}{*}{\cmarkn} & 
    \multirow{2}{*}{-} & 
    \multirow{2}{*}{R} & 
    \multirow{2}{*}{C4.5} & 
    \multirow{2}{*}{\cite{alexa1m,safebrowsinggoogle,malwaredomains}} & 
    \multirow{2}{*}{10K} & Considers only spec-\\
    & & & & & & & & & & & & ific types of DNs \\
    \hline

    \multirow{2}{*}{\cite{zhao2019malicious}} & 
    \multirow{2}{*}{\xmark} & 
    \multirow{2}{*}{\cmarkn} & 
    \multirow{2}{*}{\xmark} & 
    \multirow{2}{*}{\xmark} & 
    \multirow{2}{*}{\xmark} & 
    \multirow{2}{*}{\xmark} & 
    \multirow{2}{*}{-} & 
    \multirow{2}{*}{R} & 
    \multirow{2}{*}{Heuristic} & 
    \cite{alexa1m,spamlist,dgadata}, & 
    \multirow{2}{*}{21K} & 
    Features only\\
    & & & & & & & & & & \cite{conficker,fastfluxtracker} & & based on N-Gram\\
    \hline
    
    \multirow{2}{*}{\cite{Notosantonakakis2010notos}} & 
    \multirow{2}{*}{\xmark} & 
    \multirow{2}{*}{\cmarkn} & 
    \multirow{2}{*}{\cmarkn} & 
    \multirow{2}{*}{\xmark} & 
    \multirow{2}{*}{\xmark} & 
    \multirow{2}{*}{\cmarkn} & 
    \multirow{2}{*}{-} & 
    \multirow{2}{*}{R} & 
    \multirow{2}{*}{LBS} & 
    \multirow{2}{*}{\cite{alexa1m,malwaredomainlist,zeustracker}} & 
    \multirow{2}{*}{1.4M} & 
    Needs a lot of  \\
    & & & & & & & & & & & & history \\
    \hline
    
    \cite{Exposurebilge2011exposure}+ & \multirow{2}{*}{\xmark} & 
    \multirow{2}{*}{\cmarkn} & 
    \multirow{2}{*}{\cmarkn} & 
    \multirow{2}{*}{\xmark} & 
    \multirow{2}{*}{\xmark} & 
    \multirow{2}{*}{\cmarkn} & 
    \multirow{2}{*}{-} & 
    \multirow{2}{*}{R} & 
    \multirow{2}{*}{C4.5} & 
    \cite{alexa1m,conficker,malwaredomainlist}, & 
    \multirow{2}{*}{4.8M} & 
    Rely on passive \\
    \cite{Exposurebilge2014exposure}& & & & & & & & & &    \cite{malwaredomainsLD,phishtank,zeustracker} & & RDNS traffic \\
    \hline

    \multirow{2}{*}{\cite{Kopisantonakakis2011detecting}} & 
    \multirow{2}{*}{\xmark} & 
    \multirow{2}{*}{\xmark} & 
    \multirow{2}{*}{\cmarkn} & 
    \multirow{2}{*}{\xmark} & 
    \multirow{2}{*}{\xmark} & 
    \multirow{2}{*}{\cmarkn} & 
    \multirow{2}{*}{-} & 
    \multirow{2}{*}{R} & 
    RF, NB, KNN, & 
    \multirow{2}{*}{\cite{alexa1m,malwaredomainsLD,zeustracker}} & 
    \multirow{2}{*}{100K} & Scaling to large n/w\\
    & & & & & & & & & SVM, RC &  & & and real-time system \\
    \hline
    
    \multirow{2}{*}{\cite{stevanovic2015ground}} & 
    \multirow{2}{*}{\xmark} & 
    \multirow{2}{*}{{\xmark}} & 
    \multirow{2}{*}{{\cmarkn}} & 
    \multirow{2}{*}{\cmarkn} & 
    \multirow{2}{*}{\xmark} & 
    \multirow{2}{*}{\cmarkn} & 
    \multirow{2}{*}{-} & 
    \multirow{2}{*}{L} & 
    \multirow{2}{*}{K-Means} &  
    \multirow{2}{*}{ISP Denmark} & 
    \multirow{2}{*}{\xmark} & 
    Focus on ground- \\
    & & & & & & & & & & & & truth labeling \\
    \hline

    \multirow{1}{*}{\cite{khalil2016discovering}} & 
    \multirow{1}{*}{\xmark} & 
    \multirow{1}{*}{\xmark} & 
    \multirow{1}{*}{\xmark} & 
    \multirow{1}{*}{\cmarkn} & 
    \multirow{1}{*}{\xmark} & 
    \multirow{1}{*}{\xmark} & 
    \multirow{1}{*}{-} & 
    \multirow{1}{*}{-} & 
    \multirow{1}{*}{Heuristic} & 
    \multirow{1}{*}{\cite{alexa1m,dnsdb}} & 
    \multirow{1}{*}{54K} & 
    Based on intuition\\
    \hline

    \multirow{1}{*}{\cite{Segugiorahbarinia2016efficient}} & 
    \multirow{1}{*}{\xmark} & 
    \multirow{1}{*}{\xmark} & 
    \multirow{1}{*}{\xmark} & 
    \multirow{1}{*}{\cmarkn} & 
    \multirow{1}{*}{\xmark} & 
    \multirow{1}{*}{\xmark} & 
    \multirow{1}{*}{-} & 
    \multirow{1}{*}{-} & 
    \multirow{1}{*}{Heuristic} & 
    \multirow{1}{*}{\cite{alexa1m,malwaredomainlist}} & 
    \multirow{1}{*}{10M} & Slows down n/w\\
    \hline
    
    \multirow{1}{*}{\cite{tran2017dns}} & 
    \multirow{1}{*}{\xmark} & 
    \multirow{1}{*}{{\xmark}} & 
    \multirow{1}{*}{{\xmark}} & 
    \multirow{1}{*}{\cmarkn} & 
    \multirow{1}{*}{\xmark} & 
    \multirow{1}{*}{\xmark} & 
    \multirow{1}{*}{-} & 
    \multirow{1}{*}{-} & 
    \multirow{1}{*}{Heuristic} &  
    \multirow{1}{*}{\cite{alexa1m,dgadata,malwaredomains}} & \multirow{1}{*}{\xmark} & 
    \multirow{1}{*}{Based on intuition} \\
    \hline

    \multirow{2}{*}{\cite{liu2018imbalanced}} & 
    \multirow{2}{*}{\xmark} & 
    \multirow{2}{*}{\cmarkn} & 
    \multirow{2}{*}{\cmarkn} & 
    \multirow{2}{*}{\xmark} & 
    \multirow{2}{*}{\xmark} & 
    \multirow{2}{*}{\xmark} & 
    \multirow{2}{*}{-} & 
    \multirow{2}{*}{DI} & 
    EEA, & 
    \cite{alexa1m,cybercrimetracker}, & 
    \multirow{2}{*}{10K} & 
    Focus only on data \\
    & & & & & & & & & HAC\_EEA & \cite{hostsfile,malwaredomainsLD} & & imbalance\\
    \hline

    \multirow{2}{*}{\cite{wang2020malicious}} & \multirow{2}{*}{\xmark} & 
    \multirow{2}{*}{\cmarkn} & 
    \multirow{2}{*}{\cmarkn} & 
    \multirow{2}{*}{\xmark} & 
    \multirow{2}{*}{\xmark} & 
    \multirow{2}{*}{\cmarkn} & 
    \multirow{2}{*}{-} & 
    \multirow{2}{*}{DI} & 
    CatBoost, SVM & 
    \cite{alexa1m,Umbrella1m,safebrowsinggoogle}, & 
    \multirow{2}{*}{16K} & \multirow{2}{*}{Oversampling}\\
    & & & & & & & & & GBDT, XGBoost &\cite{malwaredomainlist,Mcafee,whois} & & \\
    \hline
    
    \multirow{2}{*}{\cite{shi2018malicious}} & 
    \multirow{2}{*}{\xmark} & 
    \multirow{2}{*}{\cmarkn} & 
    \multirow{2}{*}{\cmarkn} & 
    \multirow{2}{*}{\xmark} & 
    \multirow{2}{*}{\xmark} & 
    \multirow{2}{*}{\cmarkn} & 
    \multirow{2}{*}{APT} & 
    \multirow{2}{*}{EI} &
    ELM, LR, SVM,& 
    \cite{alexa1m,malwaredomainlist}, & 
    \multirow{2}{*}{40K} & 
    Only for \\
    & & & & & & & & & CART, BPNN & \cite{phishtank,whois} &  & 
    targeted attacks \\
    \hline
    
    \multirow{3}{*}{Ours} & 
    \multirow{3}{*}{Eth} & 
    \multirow{3}{*}{\cmarkn} & 
    \multirow{3}{*}{\cmarkn} & 
    \multirow{3}{*}{\cmarkn} & 
    \multirow{3}{*}{\cmarkn} & 
    \multirow{3}{*}{\xmark} & 
    \multirow{3}{*}{-} & 
    \multirow{3}{*}{-} & 
    K-Means, 11& 
    \cite{Umbrella1m,cryptoscam,dga}, & 
    \multirow{3}{*}{335M} & 
    \multirow{3}{*}{-}  \\
    & & & & & & & & & ML Algos &  \cite{ethereumDark,malwaredomains}, & & \\
    & & & & & & & & &  using TPOT & \cite{openphish,phishtank} & & \\
    \hline

\end{tabular}
\end{small}
\begin{tablenotes}
    \begin{small} 
        \item $\bullet$ $^{B/C}$ Blockchain, $^{Eth}$ Ethereum Blockchain data, $\bullet$ \textbf{Features}: $^{N}$ DN String based, $^{Q}$ DNS Query based, $^{G}$ DNS Graph based, $^{T}$ Temporal aspect based, $^{O}$ Other, $^\text{\xmark}$ particular feature not used, $\bullet$ \textbf{Detection Of}: $^{AG}$ Algorithmic Generated Names, $^{BT}$ Botnet, $^{FF}$ Fast-Flux, 
        $^{APT}$ Advance Persistent Threats, $^-$ no specific mention but targets DNs in general, $\bullet$ \textbf{Tackles}: $^{R}$ Reputation, $^{L}$ Ground Truth Labeling, $^{DI}$ Data Imbalance, $^{EI}$ Efficiency Improvement,  $^-$ no specific mention, $\bullet$ \textbf{ML Algo}: $^{GB}$ Gradient Boosting, $^{SVM}$ Support Vector Machine, $^{RF}$ Random Forest, $^{KNN}$ K-Nearest Neighbors, $^{NB}$ Naive Bayes, $^{BC}$ Bayesian Classifier, $^{LBS}$ Logit-Boost Strategy, $^{RC}$ Random Committee, $^{EEA}$ EasyEnsemble Algorithm, $^{ELM}$ Extreme Learning Machine, $^{GBDT}$ Gradient Boosting Decision Tree, $^{XGBoost}$ eXtreme Gradient Boosting, $^{LR}$ Logistic Regression, $^{BPNN}$ Back Propagation Neural Networks, $\bullet$ \textbf{Dataset Size}: $^\text{\xmark}$ no mention, $\bullet$ \textbf{Issues}: $^{DGA}$: Domain Generation Algorithm, $^{RDNS}$ Recursive DN System, $^{n/w}$: network, $^{IP}$: IP address
    \end{small}
\end{tablenotes}
\end{threeparttable}
\end{table*}

\begin{table*}
  \caption{Hyperparameters and Performance in related studies}
  \label{table:hy}
  \centering
  \begin{threeparttable}
  \begin{footnotesize} 
    \begin{tabular}{|c|c|cc|ccccccc|}
        \hline
         \multirow{2}{*}{\#} & \multirow{2}{*}{ML Algo.} & \multicolumn{2}{c|}{Hyperparameters} & \multicolumn{7}{c|}{Reported Performance (\%)} \\\cline{3-11}
          & & Used & Best & Accuracy & Precision & Recall & F1 & TPR & FPR & Other\\
         \hline
         
         \multirow{6}{*}{\cite{anand2020ensemble}} & \textbf{C5.0} & - & -& \textbf{97.04} & - & - & - & - & - & \textbf{94.08}$^{K}$ \\ \cline{2-11}
          & GBM & - & - &  96.44 & - & - & - & - & - & 92.87$^{K}$ \\ \cline{2-11}
          & RF & - & - & 96.63 & - & - & - & - & - & 93.26$^{K}$ \\ \cline{2-11}
          & CART & - & - & 96.27 & - & - & - & - & - & 92.55$^{K}$ \\ \cline{2-11}
          & SVM & - & - & 96.71 & - & - & - & - & - & 93.41$^{K}$ \\ \cline{2-11}
          & KNN & - & - &  94.99 & - & - & - & - & - & 89.99$^{K}$\\
         \hline
         
         \cite{Notosantonakakis2010notos} & LBS & - & - & - & - & - & - & 96.80 & 0.38 & -  \\ 
         \hline
         
         \multirow{6}{*}{\cite{Kopisantonakakis2011detecting}$\dagger$} & \textbf{RF} & - & - & - & - & - & - & \textbf{98.40} & \textbf{0.50} & - \\ \cline{2-11}
          & NB & - & - & - & - & - & - & - & - & -  \\ \cline{2-11}
          & KNN & - & - & - & - & - & - & - & - & -  \\ \cline{2-11}
          & SVM & - & - & - & - & - & - & - & - & -  \\ \cline{2-11}
          & RC & - & - & - & - & - & - & - & - & -  \\
         \hline
         
         \cite{Exposurebilge2011exposure, Exposure-bilge2014exposure} & C4.5 & - & - & - & - & - & - & - & 7.90 & 98.00$^{DR}$\\
         \hline
         
         \multirow{7}{*}{\cite{sai2020DGA}} & KNN & - & - & 96.50 & - & - & - & - & - & 95.32$^{K}$\\ \cline{2-11}
         & GBM & - & - & 97.89 & - & - & - &- & - &  97.19$^{K}$  \\ \cline{2-11}
         & C4.5 & - & - & 97.56 & - & - & - & - & - & 96.75$^{K}$ \\ \cline{2-11}
         & \textbf{C5.0} & - & - & \textbf{98.08} & - & - & - & - & - & \textbf{97.44}$^{K}$  \\ \cline{2-11}
         & CART & - & - & 82.30 & - & - & - &- & - &  76.40$^{K}$ \\ \cline{2-11}
         & RF & - & - & 98.02 & - & - & - & - & - & 97.37$^{K}$ \\  \cline{2-11}
         & NB & - & - & 77.58 & - & - & - & - & - & 70.11$^{K}$ \\ 
         \hline 
         
         \multirow{4}{*}{\cite{hoang2018botnet}} & KNN & k $\in$ [1, 20] & 15 & 87.70 & - & - & 86.70 & 80.40 & 5.00 & 94.10$^{DR}$ \\ \cline{2-11}
         & C4.5 & - & - & 87.60 & - & - & 86.70 & 80.90 & 5.70 & 93.40$^{DR}$ \\  \cline{2-11}
         & \textbf{RF} & T $\in$ [5, 50] & 27 & \textbf{88.10} & - & - & \textbf{87.20} & \textbf{80.90} & \textbf{4.80} & \textbf{94.40}$^{DR}$  \\  \cline{2-11}
         & NB & - & - & 86.00 & - & - & 86.40 & 89.10 & 17.10 & 83.90$^{DR}$ \\
         \hline
         
         \cite{khalil2016discovering} & Heuristic & - & - & - & - & - & - & 95.00 & 0.50 & - \\
         \hline
         
         \multirow{2}{*}{\cite{liu2018imbalanced}} & \textbf{HAC\_EEA} & - & - &  \textbf{96.32} & \textbf{95.17} & - & \textbf{95.74} & - & - & - \\ \cline{2-11}
         & EEA & - & - & 93.58 & 93.38 & - & 93.47 & - & - & - \\
         \hline
         
         \cite{messabi2018malware} & C4.5 & - & - & - & 85.20 & 69.90 & 76.80 & 69.90 & 13.80 & 77.50$^{DR}$\\
         \hline
         
         \cite{perdisci2009detecting} & C4.5 & - & - & - & - & - & - & - & 0.70 & 99.30$^{DR}$ \\
         \hline
         
         \cite{FluxBusterperdisci2012early} & C4.5 & - & - & - & - & - & - & - & 0.15 & 99.30$^{DR}$ \\
         \hline
         
         \cite{Segugiorahbarinia2016efficient} & Heuristic & - & - & - & - & - & - & 95.00 & 0.10 & -\\
         \hline
         
         \cite{selvi2019detection} & RF & - & - & 98.73 & - & - & - & - & - & 97.47$^{K}$ \\
         \hline
         
         \multirow{5}{*}{\cite{shi2018malicious}} & \textbf{ELM} & - & - & \textbf{96.28} & - & - & - & - & - & - \\ \cline{2-11}
         & LR & - & - & 91.95 & - & - & - & - & - & - \\ \cline{2-11}
         & CART & - & - & 91.83 & - & - & - & - & - & - \\ \cline{2-11}
         & BPNN & - & - & 95.82 & - & - & - & - & - & - \\ \cline{2-11}
         & SVM & - & - & 94.70 & - & - & - & - & - & - \\
         \hline
        
         \multirow{3}{*}{\cite{stalmans2011framework}} & C5.0$\ddagger$ & - & - & - & - & - & - & - & - & - \\ \cline{2-11}
         & BC & - & - & 85.00 & - & - & - & 81.00 & 11.00 & - \\ \cline{2-11}
         & Naive-BC & - & - & 87.00 & - & - & - & 82.00 & 8.00 & -  \\
         \hline
         
         \cite{stevanovic2015ground} & K-Means & - & - & 73.48 & 72.41 & - & - & 52.69 & 13.02 & - \\
         \hline
         
         \cite{tran2017dns} & Heuristic & - & - & 98.30 & 99.10 & 98.60 & - & - & - & - \\
         \hline
         
         \cite{truong2020detecting} & RF & - & - & 98.79 & 99.86 & - & - & 94.52 & 0.13 & 2.19$^{FNR}$ \\
         \hline
         
         \multirow{4}{*}{\cite{wang2020malicious}} & \textbf{CatBoost} 
         & - & & \textbf{98.42} & \textbf{99.26} & - & \textbf{98.38} & - & - & 52.518s$^{rTime}$\\ \cline{2-11}
         & SVM & - & - & 94.84 & 95.08 & - & 96.41 & - & - & 174.564s$^{rTime}$ \\ \cline{2-11}
         & GBDT & - & - & 94.78 & 95.07 & - & 95.07 & - & - & 94.912s$^{rTime}$\\ \cline{2-11}
         & XGBoost & - & - & 94.81 & 95.07 & - & 96.39 & - & - & 89.074s$^{rTime}$ \\
         \hline
         
         \cite{zhao2019malicious} & Heuristic & - & - & 94.04 & - & - & - & - & 6.14 & 7.42$^{FNR}$\\
         \hline
         
    \end{tabular}
    \end{footnotesize}
    \begin{tablenotes}
    \begin{scriptsize}
        \item $^{TPR}$ True Positive Rate, $^{FPR}$ False Positive Rate, $^{FNR}$ False Negative Rate, $^{TNR}$ True Negative Rate, $^{DR}$ Detection Rate, $^{rTime}$ Running Time, $^{TO}$ Time Overhead, $^{K}$ Kappa Comparison, $^-$ no mention, $^\dagger$ only mentions for the best case, $^\ddagger$ no results shown
    \end{scriptsize}
    \end{tablenotes}
  \end{threeparttable}
\end{table*}

\vspace{-0.3cm}
\section{Methodology}\label{sec:method}

We follow a standard ML pipeline that includes the following main processing steps: data collection, data pre-processing (cleaning, enrichment, and ground truth labeling), feature engineering (feature extraction and selection), ML algorithm (supervised and unsupervised), and evaluation. This section mainly focuses on the data pre-processing, feature engineering, and application of ML algorithm steps.

\vspace{-0.4cm}
\subsection{Data Collection and Pre-processing}

In blockchains, meta-information associated with dApps and BCWs is highly limited. On top, in Ethereum, extraction of DNs associated with dApps and BCWs is not easy. Etherscan and other blockchain explorers provide APIs to extract transaction data of accounts present in Ethereum.  
Further, very few accounts are labeled to be involved in malicious activity. Due to such limitations, we evaluate our approach using DN data available from multiple public sources such as~\cite{Umbrella1m,cryptoscam,dga,ethereumDark,malwaredomains,openphish,phishtank}. Note that~\cite{cryptoscam,ethereumDark} provides a limited list of malicious DNs present in the Ethereum Blockchain, while~\cite{Umbrella1m} provides us with the DNS records for the DNs over time.~\cite{dga,malwaredomains,openphish,phishtank} provides us the ground truth labels for the malicious DNs. 

A DNS query record of~\cite{Umbrella1m} typically contains 105 fields per record. However, we only consider those fields that are useful for us. These include \emph{query} (or the DN), \emph{query type}, \emph{response type}, \emph{response name}, \emph{TTL} (Time-to-Live), \emph{timestamp}, \emph{RTT} (Round-Trip-Time), \emph{IP Address} (IPv4 and IPv6), \emph{country}, \emph{IP prefix}, \emph{CNAME} (Canonical Name), \emph{DNAME} (Delegation Name), \emph{MX Address} (Mail Exchange), \emph{NS Address} (Name Server), and \emph{TXT Records}. We clean this data by removing the records in which information about either TLD (top-level domain extracted from the query/DN) or timestamp is missing. We also remove the strings like `www.' and `http' from the DN of the DNS query. After cleaning, we enrich our data by labeling each DN as malicious or benign. The labels or the ground truth information about the DNs is extracted from~\cite{cryptoscam,dga,ethereumDark,malwaredomains,openphish,phishtank}.

In the pre-processing stage, we segment the DNS record data based on different temporal granularities that range from hours to months to identify behavioral changes. Due to computational resources available to us, we currently only validate our approach on two temporal granularities ($T_G$): 2 hour ($2H$) and All ($ALL$). In a $2H$ temporal granularity, the entire data $D$ is segmented into multiple data segments of 2 hours each. For example, if the entire data is of 1 month (duration $T$ hours), we have $\approx$1$\times$30$\times$12 data segments. In $ALL$ granularity we use entire data instead of segmenting it. Note that a data segment in the $2H$ granularity is not a static snapshot, but the features present in the feature vector are both temporal and static. Here, 2-hour granularity provides fine-grain results and is able to capture highly dynamic behavior. 

\vspace{-0.3cm}
\subsection{Feature Engineering}

We use temporal (i.e., time-series based) and non-temporal (i.e., non-time series based) features to understand the actual behavior of a DN before we classify it as malicious or benign. We extract all the non-temporal features (described next) from the DNs using string-based analysis and DNS query/response information given in DNS traffic records during feature engineering. The temporal features (such as \emph{burst}, \emph{degree}, and \emph{diameter} also described next) are the time-dependent features reflecting the changes. The analysis of the DNS traffic data enables the identification of such features. 

Thus, our set of features ($F$) is based on both non-temporal (that comprise of string-based features ($F_s$) and DNS query-based ($F_q$) features) and temporal features (that comprise of burst-based features ($F_b$) and temporal graph-based ($F_g$) features). In details, these features are:

\begin{itemize}
    \item \textbf{Non-Temporal features:}
    The string-based features ($F_s$) and DNS query-based ($F_q$) features are mainly extracted using the lexical and statistical analysis. These features are directly derived from the DN string and DNS traffic records.
    
    \begin{itemize}
        \item \textbf{String-based features ($F_s$):} are useful to detect the malicious DN that are algorithmically generated by a Domain Generation Algorithm (DGA) or follow specific string patterns. These patterns are generally identified after performing the lexical and statistical analysis. From the analysis, we identify features such as the length of the DN, number of digits in the DN, number of unique digits in the DN, number of characters in the DN, number of unique characters in the DN, number of symbols in the DN, number of vowels in the DN, number of consonants in the DN, number of unique  alphanumeric characters, and the ratio between the number of unique alphanumeric characters and the total length of the DN. Note that we do not use the N-gram-based features due to the limited computation power we have. 
        
       \item \textbf{DNS Query-based features ($F_q$):} are useful for identifying malicious DNs as there are significant differences between DNS traffic footprints of a malicious DN and a benign DN. With respect to static features, we extract features such as the ratio between each of the unique \emph{query types}, \emph{response type}, \emph{response name}, \emph{counties}, \emph{IP prefix}, \emph{CNAME}, \emph{DNAME}, \emph{MX Address}, \emph{NS Address}, \emph{TXT Records}, \emph{IPv4}, \emph{IPv6} and respective total count value from the specific DNS query records of a DN. Apart from these ratios, other features includes the number of unique \emph{TTLs}, minimum TTL, maximum TTL, average TTL, standard deviation in TTL, number of unique \emph{RTTs}, minimum RTT, maximum RTT, average RTT, standard deviation in RTT, total number of unique IP addresses (sum of number of IPv4  and IPv6 addresses), number of unique \emph{IPv4 addresses}, and number of unique \emph{IPv6 addresses} associated with a DN.  
    \end{itemize}
\end{itemize}

\begin{figure}
\vspace{-0.3cm}
\centering
  \includegraphics[width=0.95\columnwidth]{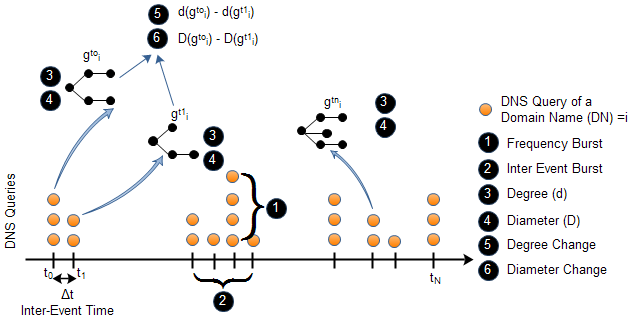}
\caption{Concept of temporal features}
\label{figure1}
\vspace{-0.3cm}
\end{figure}

The DNS traffic data is temporal data. Thus, we extract a temporal graph ($g^t_i(V^t,E^t)$) at time $t$ representing the connections between the DN ($i$) and the associated IPs. Let the set of such temporal graphs be $G$. Here, $V^t$ is a set of nodes such that a node represents a type $\in$ \emph{\{DN, IPv4 address, IPv6 address, NS address, IP prefix\}} and $E^t$ represents a set of link/edge between the nodes in $V^t$. At each $t$, we create temporal graphs and extract properties such as \emph{Diameter}, \emph{Degree}, and their \emph{changes} (these features are described next). Note that these features are derived from the time series analysis of the DNS traffic record. The overall notion of the temporal features is illustrated in Figure~\ref{figure1}. Here, the figure represents the DNS queries for a particular DN. These queries are fired at different times where the difference between the two consecutive timestamps is represented by inter-event time $\Delta t$. At each time instance, there could be multiple DNS queries (query frequency). Further, the DNS queries could be consecutive and span over time. A $g^t_i$ is computed using all the DNS queries at a time $t$ and has properties such as degree and diameter. Next, we explain the different temporal features (query frequency burst, query inter-event burst, degree of DNS query, and diameter of DNS query) that we derive.   

\begin{itemize}    
    \item \textbf{Burst-based features} ($F_b$): A burst is defined as a temporal non-homogeneous sequence of DNS queries. We consider two types of bursts (query frequency burst (fB) and query inter-event burst (iB)). Both bursts are derived from the temporal analysis of the DNS traffic data. 
    
    \begin{itemize}
        \item \textbf{Query frequency burst} (fB): represents the number of DNS queries at a particular time instance. If the number of queries at a particular time is above a threshold (i.e., 80\% of the maximum number of queries of a DN during a time frame), then it considers as a candidate of frequency burst. From the analysis of candidates, we formulate burst features such as the number of frequency bursts of a DN, maximum size of frequency bursts of a DN, average frequency burst of a DN.
    
        \item \textbf{Query Inter-Event Burst} (iB): refers to a continuous sequence of DNS queries where the inter-event time is very small. The DNS queries that were continuously made, up to a certain threshold (i.e., more than 80\%), their DNs are considered as a candidate of the inter-event burst. From the analysis of candidates, we define features such as the number of inter-event bursts of a DN and the maximum size of an inter-event burst of a DN.
    \end{itemize}
\end{itemize}  

\begin{figure*}
\vspace{-0.3cm}
\centering
  \includegraphics[width=0.85\textwidth]{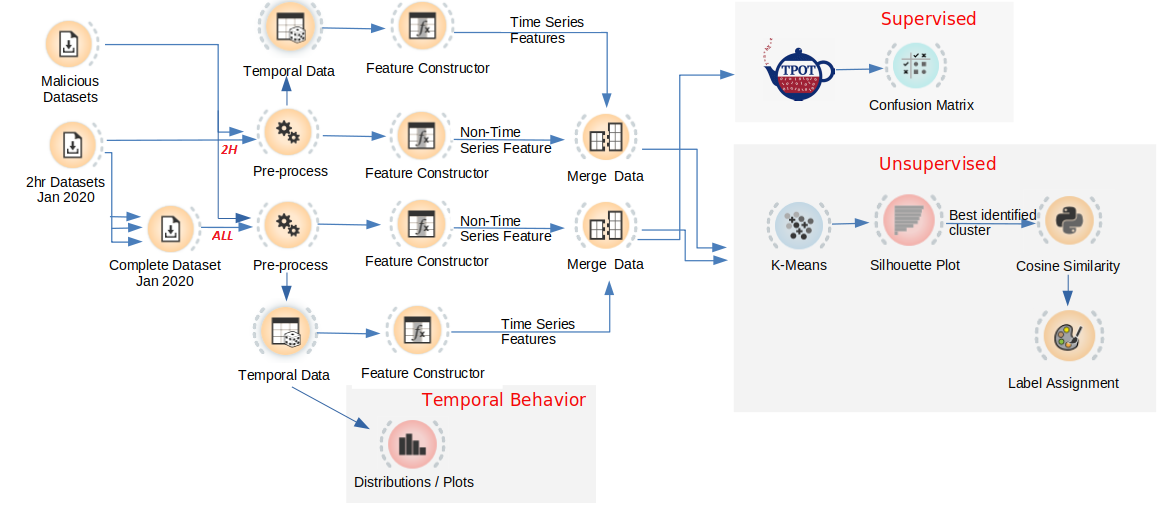}
\caption{Our Methodology}
\label{figure10}
\vspace{-0.3cm}
\end{figure*}

\begin{itemize}
    \item \textbf{DNS graph-based features} ($F_g$): These features are directly derived from the temporal graph ($g^t_i$) of a DN ($i$). These are useful to identify the malicious DNs based on the association and connection between the elements of the DNS network.
 
    \begin{itemize}
        \item \textbf{Degree} ($d$): In $g^t_i$, degree represents the number of edges that a DN ($i$) has. It is helpful to understand whether the malicious DN changes their associated IPs frequently.
        Here, we consider features such as the ratio between the number of times degree changes and total time instance the DN appeared and the ratio between total degree change and total time instance. Assume that a DN's change in degree over time is represented by a vector (2, 0, 0, 2, 1, 2), then the two above-mentioned feature values are 4/6 and 7/6, respectively. Apart from these two ratios, we also consider the maximum degree and minimum degree of a DN as degree features. 
        
        \item \textbf{Diameter} ($D$): is useful to detect malicious DNs based on their association with others. It represents the largest shortest path of the component in which the DN exists. From the analysis of this component, we formulate features such as the maximum diameter of a DN, the minimum diameter of a DN, and the number of times diameter changes during the time frame. 
    \end{itemize}
\end{itemize}
  
After the feature extraction, as the features might be correlated and may not reflect much importance, we select the features based on the Pearson correlation. Here, we remove those features for which the Pearson correlation value is high. 

\vspace{-0.3cm}
\subsection{ML Algorithm}

After the feature engineering process, we apply supervised and unsupervised ML algorithms to different data segments generated at different temporal granularities. This helps us analyze the behavior of DNs and identify the DNs which show persistent suspicious behavior over time. Most of the ML algorithms used in the state-of-the-art approaches are supervised ML algorithms. Thus, to validate and test which supervised algorithm performs best in classifying malicious DNs, we use AutoML tool called TPOT~\cite{olson2016tpot}. Here, we configure TPOT to use the supervised algorithms and their set of hyperparamters as reported by state-of-the-art algorithms as well validate other sets of hyperparameters because it may enhance the overall results. Note that we use default hyperparameters as set by the used python library in the case where the state-of-the-art approaches do not report hyperparameters. We report balanced-accuracy, Precision, Recall, and F1-score for the supervised case. We apply supervised algorithms only on \emph{ALL} data granularity. This reduces our search space to stay within the limits of the computing power available to us.

Further, we also use unsupervised ML algorithms to validate our approach. Here, to test, we only use the K-Means clustering algorithm due to the availability of computing resources and use hyperparameter K$\in$[30, 100] (for \emph{ALL} granularity) and K$\in$[7, 24] (for \emph{2H} granularity). Our choice (range on K) is based on the size of data. For the best number of clusters identified using silhouette score, we select the cluster where most malicious DNs are. We then identify the benign account within the selected cluster that shows high similarity (similarity$\rightarrow$1) with the malicious account. Here, to calculate the similarity, we chose cosine similarity as it is widely adopted. We acknowledge that there are other types of similarity measures, and with the use of different similarity measures, the results might change. But for this work, we chose cosine similarity simply because it is widely used and accepted. In the case of \emph{2H}, we repeat the same process for each data segment and determine the behavior of DNs based on the probability, which is calculated as the fraction of the number of times a DN shows malicious behavior over the total number of times the DN occurs.

\begin{figure*}
\vspace{-0.3cm}
    \centering
    \subfloat[Distribution of Query Frequency of individual DNs]{
        \includegraphics[width=0.9\columnwidth]{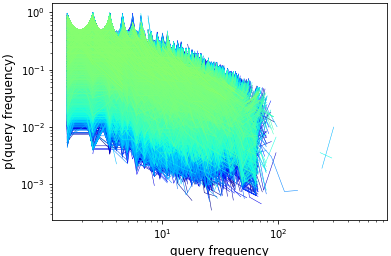}\label{fig:queryFreqInd}
    }
    \subfloat[Distribution of Query Frequency of all DNs combined]{
        \includegraphics[width=0.9\columnwidth]{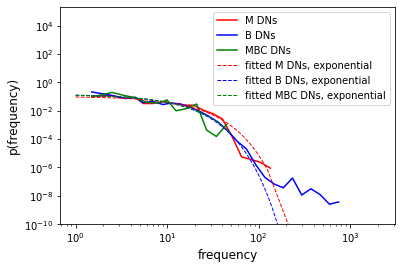}
        \label{fig:queryFreq}
    }
    \caption{Distribution of Query Frequency}%
    \label{figure2}
    \vspace{-0.3cm}
\end{figure*}

In summary, our methodology is illustrated and summarized in Figure~\ref{figure10}. It shows the standard ML pipeline that includes key processing steps: pre-process, validation of temporal behavior, feature constructor, and ML algorithms. The collected dataset is organized into two temporal granularities: 2H and ALL. During pre-processing, data is cleaned by removing the noisy records and labeled using the malicious datasets. The time-series data is extracted from the desired fields and processed to generate different features. Next, to understand temporal behavior that exists, we validate it on different data segments representing different temporal granularities. Time-series and non-time series features are constructed from the respective datasets. Finally, a complete feature vector is formed by merging both time-series and non time-series features and is passed to the ML algorithms.

\vspace{-0.3cm}
\section{Evaluation and Results}\label{sec:results}

We evaluate our approach on the publicly available Cisco Umbrella top 1 million dataset~\cite{Umbrella1m} for January 2020. All evaluations are performed using Python v3.8.3 within the Python environment with supporting libraries such as pandas v1.1.3, numpy v1.18.5, pickle v4.0, matplotlib v3.2.2, and tldextract v2.2.3. With respect to computational resources, we test our approach on a machine with Intel(R) Core(TM) i7-9700 CPU@3.00 GHz with 64.00 GB RAM.

\vspace{-0.4cm}
\subsection{Dataset}

\emph{Cisco Umbrella top 1 million dataset}~\cite{Umbrella1m} for the month of January 2020 contains 361 sub-datasets of $\approx$2 hours each. For January 2020, there are $\approx$335 Million DNS queries and out of these, only $\approx$1.77 Million are distinct queries because dataset has multiple entries for a query in dataset with different DNS records. Note that we consider only January 2020 data due to the limited computing power available to us. For the malicious tags/labels, we use DGA dataset \cite{dga}, Phishtank \cite{phishtank}, Openphish \cite{openphish}, Cryptoscam blacklist \cite{cryptoscam}, Ethereum wallet dark list \cite{ethereumDark} and Malware DN list \cite{malwaredomains}. DGA dataset has $\approx$88 Million malicious DNs on 07$^{th}$ January 2019. The total number of malicious DNs extracted from sources other than DGA is 31985. Out of these, 14706 are unique TLDs (top-level domains). In the complete dataset, there are only 87557 malicious DNs, out of which 41991 DNs are present in different blockchains such as Ethereum. Note that these 41991 malicious DNs are tagged as of 30$^{st}$ October 2020. We also use a new list of unique 2479 malicious blockchain DNs~\cite{cryptoscam, ethereumDark} as of 7$^{th}$ April 2021 and used it to validate our results. 

\vspace{-0.3cm}
\subsection{Validation of Temporal aspects and importance of a feature}

For validation of temporal aspects, on the complete dataset, we analyze the behavior of temporal features present in malicious DNs, benign DNs, and malicious DNs in a blockchain. Although more refined, the \emph{2H} granularity has very limited data and variation. In the following discussion, we represent malicious DNs present in a Blockchain as \textit{MBC}.

To validate the existence of the \emph{query frequency bursts}, we study the distribution of the number of queries made (or the query frequency) for each DNs (cf. Figure~\ref{fig:queryFreqInd} where a different color represents a different DN). We also study the distribution of query frequencies per class computed considering all individuals as one under that class (cf. Figure~\ref{fig:queryFreq}). Figure~\ref{fig:queryFreqInd} shows a bursty \emph{query frequency} where very few DNs have a high query frequency (more than 100) while most of the other DNs have a low query frequency (less than 100). Similarly, when analyzing the differences between DNs those belong to benign (\textit{B}), malicious (\textit{M}), and \textit{MBC} class, we identify that the probability distribution of query frequency follows an exponential distribution with $x_{min}$=1.0 and $\lambda$=$\{0.1313, 0.0958, 0.1267\}$, respectively. Here, we observe that the distributions for \textit{B} and \textit{MBC} are relatively similar. To identify how similar they are, we use the KL Divergence (KLD). A KLD=0 signifies that the two distributions are similar, while a KLD$>$0 signifies that the two distributions diverge. In our case, between the distributions for \textit{B} and \textit{MBC} class, KLD is 6.28$\times$10$^{-4}$. From this, we infer that the two classes have differences, and query frequency bursts can be used as a feature.

\begin{figure*}
    \centering
    \subfloat[Distribution of Inter-Event Time in millisecond of individual DNs]{
        \includegraphics[width=0.9\columnwidth]{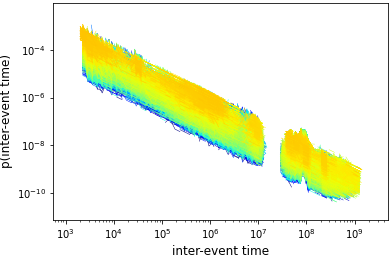}
        \label{fig:IETInd}
    }
    \subfloat[Distribution of combined Inter-Event Times in millisecond]{
        \includegraphics[width=0.9\columnwidth]{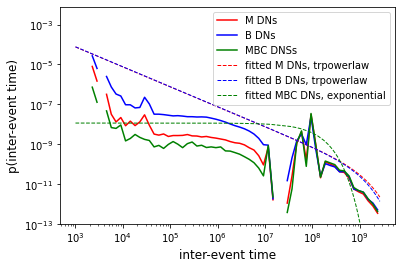}
        \label{fig:IET}
    }
    \caption{Distribution of Inter-Event Time}%
    \label{figure3}%
\end{figure*}

For an inter-event time, in Figure~\ref{figure3}, we also observe the existence of a burst. Here, Figure~\ref{fig:IETInd} shows that for a very large number of DNs the maximum inter-event time $<$10$^4$, while a very few DNs have a very high (maximum $>$10$^8$) inter-event time. Note that as we show distributions of the individual (represented by a different color) inter-event times, these distributions have different minimum and maximum values resulting in a visual disconnect (i.e., absence of inter-event times for certain values). Similarly, for the \textit{B}, \textit{M}, and \textit{MBC} cases, the probability distribution of inter-event time is shown in Figure~\ref{fig:IET}. Here, on the entire data due to diversity in $x_{min}$ and $x_{max}$ values for each DNs the studied distributions had a large Kolmogorov-Smirnov (KS) distance. Nonetheless, out of the studied distributions, truncated-powerlaw provided the best fit. For the truncated-powerlaw distribution, for the two classes (\textit{B} and \textit{M}), $x_{min}$=1000 millisecond and a cutoff parameter, $\beta$=$\frac{1}{\lambda}$ where $\lambda$=$\{1.1563$$\times$$10^{-9}, 9.6612$$\times$$10^{-9}\}$, while for \textit{MBC} exponential distribution was the best identified fit with parameters $x_{min}$=1000 and $\lambda$=$1.1505$$\times$$10^{-8}$. The whole notion of distribution fit is dependent on the sample space (more specifically, the range of `x' ($x_{min}$ and $x_{max}$) in which the distribution is studied). Choosing $x_{min}$=5000 and $x_{max}$=$10^7$, we identify that exponential distribution with parameter values $\lambda$=3.9897$\times$10$^{-7}$ for \textit{M}, $\lambda$=4.3806$\times$10$^{-7}$ for \textit{B}, and $\lambda$=4.2957$\times$10$^{-7}$ for \textit{MBC} fits the best. Thus, this feature can also be an indicator.

\begin{figure*}
\vspace{-0.3cm}
    \centering
    \subfloat[Distribution of Degree of individual DNs]{
        \includegraphics[width=0.9\columnwidth]{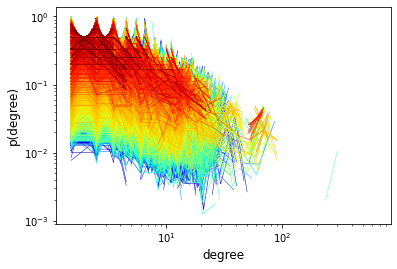}
        \label{fig:DegInd}
    }
    \subfloat[Distribution of Degree of all DNs combined]{
        \includegraphics[width=0.9\columnwidth]{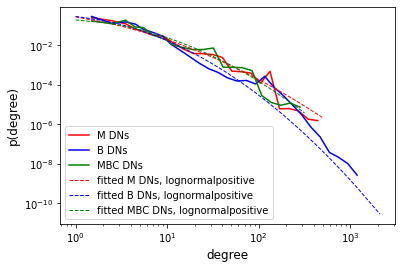}
        \label{fig:Deg}
    }
    \caption{Distribution of Degree}%
    \label{figure5}%
    \vspace{-0.3cm}
\end{figure*}

\begin{figure}
\vspace{-0.3cm}
\centering
  \includegraphics[width=0.9\columnwidth]{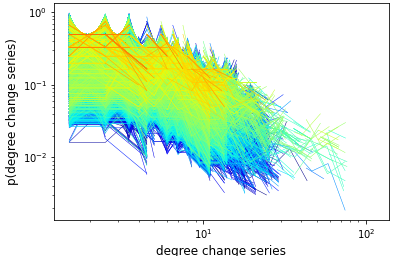}
\caption{Distribution of Degree Change of individual DNs}
\label{fig:DegChInd}
\vspace{-0.3cm}
\end{figure}

\begin{figure*}
    \centering
    \subfloat[Distribution of Count of Degree Changes of all DNs combined]{
        \includegraphics[width=0.9\columnwidth]{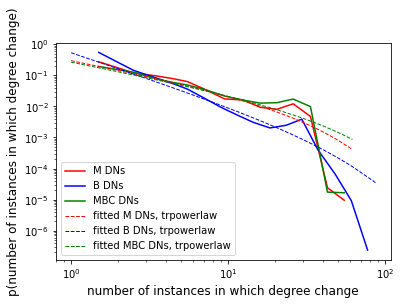}
        \label{fig:DegChCount}
    }
    \subfloat[Distribution of Sum of Degree Changes of all DNs combined]{
        \includegraphics[width=0.9\columnwidth]{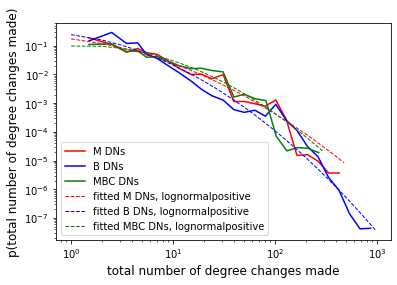}
        \label{fig:DegChSum}
    }
    \caption{Distribution of Degree Change}%
    \label{figure6}
\end{figure*}

Next, we identify the behavior of the degree present in the data by analyzing the change in degree over time. First, we analyze the degree distribution plot (cf. Figure~\ref{figure5}) and then the degree change distribution plot (cf. Figure~\ref{fig:DegChInd} and \ref{figure6}). Figure~\ref{fig:DegInd} shows the variations in the degree of each DN. Here, we observe that \emph{(i)} very few DNs have a degree $>$100 while most of the other DNs have a degree $<$10 (thus, showing a bursty behavior) and \emph{(ii)} for most of the DNs, change in the degree exists as their degree range $\in[1,100]$. The distribution plots of the DNs with respect to their types follows a positive log-normal distribution with $x_{min}$=1.0, $\mu$=$\{0.7719, 0.8106, 1.3198\}$, and $\sigma$=$\{1.6206, 1.2229, 1.3250\}$ for \textit{M}, \textit{B} and \textit{MBC} classes, respectively. Here, KL divergence between \textit{M} and \textit{B} is $0.0970$, \textit{M} and \textit{MBC} is $0.1321$, \textit{B} and \textit{MBC} is $0.07994$. Further, we also observe that the maximum degree for \textit{B} is $>$1000 while that for \textit{M} and \textit{MBC} types is $\approx$500 and $\approx$300, respectively. Similarly, Figure~\ref{fig:DegChInd} shows the distribution of the change in degree for each DN. As discussed before, a change is a difference between degree values in consecutive time instances. However, if the degree reduces, we do not consider it as a change. From Figure~\ref{fig:DegChInd}, we observe that very few DNs show a change $\geq$70 while most of DNs have a change that is $\leq$10. We also observe that no DN has a change $>$100. 

We then analyze the behavior differences between \textit{M}, \textit{B} and \textit{MBC} type of DNs in terms of the number of instances in which a degree change was made (cf. Figure~\ref{fig:DegChCount}) and the total number of such changes made (cf. Figure~\ref{fig:DegChSum}). In the case of number of instances in which a degree change happened, we find that for each DN the distribution again showed a bursty behavior. When analysing particular classes of DNs, we find that the distribution for each class (\textit{B}, \textit{M}, and \textit{MBC}) follows a truncated-powerlaw distribution with $x_{min}$=1.0, $\alpha$=$\{1.0001, 1.6838, 1.0001\}$ and $\lambda$=$\{0.0404, 0.0245, 0.0264\}$. We also find that the maximum number of degree changes made at any instance for \textit{B} is $\approx$90, while for both \textit{M} and \textit{MBC} it is $\approx$80. Similarly, in case of total number of degree changes made, we find that positive log-normal distribution fits best with $x_{min}$=1.0, $\mu$=$\{1.5286, 1.0357, 1.9800\}$, and $\sigma$=$\{1.5968, 1.3693, 1.3270\}$ for \textit{B}, \textit{M}, and \textit{MBC}, respectively. Here, KLD between the distributions identified for \textit{M} and \textit{B} is 0.09103. While that between \textit{M} and \textit{MBC} is 0.0967 and between \textit{B} and \textit{MBC} is 0.2542.

\begin{figure*}
\vspace{-0.3cm}
    \centering
    \subfloat[Variation between maximum and minimum of Diameter]{
        \includegraphics[width=0.9\columnwidth]{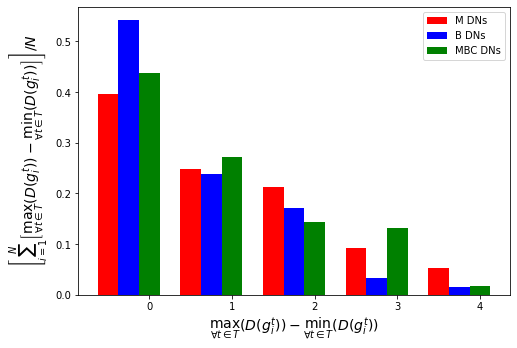} 
        \label{fig:DiaMM}
    }
    \subfloat[Distribution of Diameter Change of all DNs combined]{
        \includegraphics[width=0.9\columnwidth]{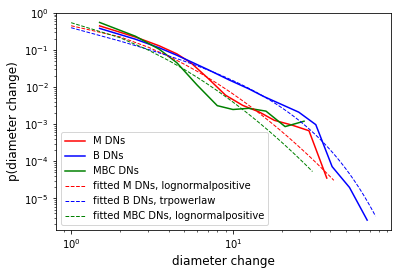} \label{fig:DiaCh}
    }
    \caption{Distribution of Diameter}
    \label{figure7}
    \vspace{-0.3cm}
\end{figure*}

Next, we identify the behavior of the diameter attribute present in the DNS traffic data. We study the $Diff$=$\max_{\forall t\in T}(D(g_i^t))-\min_{\forall t\in T}(D(g_i^t))$ associated with a DN over entire data and the probability distribution of the number of instances in which diameter changes occur for a DN. Here, $T$ is set of all times when a DNS query was made for a particular DN, $i$. Figure~\ref{fig:DiaMM} shows the probability of $Diff$ associated with different types of DNs. Here, we observe that the $\max(Diff)$ for any DN is 4, but such instances are rare. For most of DNs, there was no change in the diameter. For \textit{B} DNs $\approx$54\% DNs showed no change. We also observe the significant difference between \textit{B} and \textit{M} DNs. Similarly, the analysis of probability distribution of number of instances in which diameter changes (cf. Figure~\ref{fig:DiaCh}) identifies the positive log-normal distribution fits \textit{M} and \textit{MBC} DNs, while truncated-power-law distribution fits \textit{B} DNs. The parameters of probability distribution for \textit{M} and \textit{MBC} DNs are $x_{min}$=1.0, $\mu$=$\{0.3208, 1.3323$$\times$10$^{-15}\}$, and $\sigma$=$\{0.9940, 1.0115\}$, respectively. The KLD between \textit{M} and \textit{MBC} DNs is 0.0506. The distribution parameters of \textit{B} DNs are $x_{min}$=1.0, $\alpha$=1.0540, and $\beta$=$\frac{1}{\lambda}$ where $\lambda$=0.0949. The different distributions show differences between the \textit{B} and \textit{M} DNs.  

\begin{figure}
\vspace{-0.3cm}
\centering
\includegraphics[width=0.9\columnwidth]{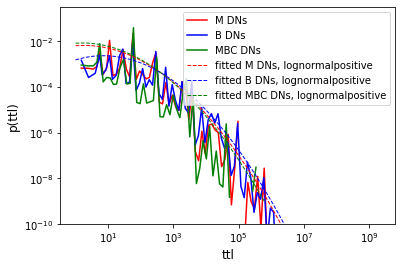}
\caption{Distribution of Time-to-Live of all DNs combined}
\label{figure8}
\vspace{-0.3cm}
\end{figure}

Apart from the above discussed temporal features, Time-to-Live (TTL) also reflects the behavioral difference between the \textit{B}, \textit{M}, and \textit{MBC} DNs. Here, we study a composite distribution of TTL of all DNs under a class (cf. Figure~\ref{figure8}). We observe that a positive log-normal distribution fits all classes of DNs (i.e., \textit{B}, \textit{M}, \textit{MBC}) with $x_{min}$=1.0, $\mu$=$\{5.9656, 6.7587, 5.6341\}$, and $\sigma$=$\{2.3297, 2.1915, 2.2833\}$, respectively. Here, we observe the KLD between the \textit{M} and \textit{B} distributions is 0.0691. While that between \textit{M} and \textit{MBC} is 0.0116 and between \textit{B} and \textit{MBC} is 0.1244. It signifies that the distributions small divergence with each other.

From the above analysis, we see that temporal aspects exist in DNs where DN associated with \textit{B} and \textit{M} differ. While those that belong to \textit{M} and \textit{MBC} show less divergence than with \textit{B}. Thus, these temporal features along with the non-temporal features play as an important role in differentiating \textit{M} and \textit{B} DNs, but not so much between \textit{M} and \textit{MBC} DNs.

\vspace{-0.4cm}
\subsection{Results}

As the feature space is large, we use Pearson correlation to identify correlated features. We find that there is a weak correlation between the features. Thus, we use all the 48 features and train the ML algorithms.

\subsubsection{Approach using Supervised Learning}

Most state-of-the-art approaches use supervised ML algorithms. Thus, we validate and test which supervised ML algorithm, along with its hyperparameters, performs best towards identifying the malicious DNs in the complete dataset. To perform such a task, we use AutoML tool called TPOT. Although other AutoML tools exist, our choice to use TPOT is based on its easy-to-use functionality. Here, we configure TPOT to use 11 different supervised ML algorithms that are used by the state-of-the-art approaches. We configure these 11 algorithms to use hyperparameters reported by the state-of-the-art and other custom hyperparameters to have more diversity. For the best identified algorithm in terms of balanced-accuracy, we also report Precision, Recall, and F1 score for the malicious class. 

TPOT reports DecisionTree (criterion=entropy, max\_depth=8, min\_samples\_leaf=5, min\_samples\_split=18, splitter=random) to achieves \emph{balanced-accuracy} of 89.53\% as the best classifier for malicious blockchain DNs. Here, other hyperparameters have default values and thus not reported here. It achieves 95.0\% \emph{Precision}, 79.16\% \emph{Recall}, and 86.0\% \emph{F1 score}. In case of detection of all malicious DNs (including malicious blockchain DNs), 
TPOT reports DecisionTree (max\_depth=10, min\_samples\_ leaf=13, min\_samples\_split=12) to achieves \emph{balanced-accuracy} of 81.52\% and as the best for detection. It achieves 91.0\% \emph{Precision}, 63.0\% \emph{Recall}, and 75.0\% \emph{F1 score}. 
Although, the Recall on the malicious class is low, the overall balanced-accuracy is high. Nonetheless, selecting with respect to the best Recall as the selection criteria, TPOT reports GaussianNB (var\_smoothing=$\times$10$^{-8}$) to achieve best \emph{Recall} of 98.58\% on the malicious class. While \emph{Precision} is 5.0\% and \emph{F1-score} is 9.0\% on the malicious class. Here, a high Recall value or the low false-negative score demonstrates that most malicious DNs are correctly identified. However, the balanced-accuracy is low (the Recall value is 1.0\% for benign class). Thus, reducing the balanced-accuracy to $\approx$50.0\%. 

\subsubsection{Approach using Unsupervised Learning}

We apply K-Means unsupervised ML algorithm on the entire dataset to detect malicious DNs (those that have high cosine similarity ($>$1-$\epsilon$ were $\epsilon$=10$^{-7}$) with malicious DNs). Using K-Means, we find optimal cluster configurations (value of K) based on the silhouette score. We check the silhouette score for different values of K$\in$[30,100] and find optimal silhouette score is 0.836 for K=50 (different silhouette scores obtained are listed in Table~\ref{table:5}). After exploring the clusters obtained for K=50, we find that cluster number 22 has 77751 malicious DNs and 1284523 benign DNs. Identifying cosine similarity for such a large number of DNs is resource expensive and the approach does not capture behavioral changes.

\begin{table}
    \caption{Silhouette Scores (S) obtained by K-Means for different K.}
    \label{table:5}
    \centering
    \begin{tabular}{|c|c|c|c|c|c|c|c|c|}
        \hline
         \textbf{K} & 30 &40& 50&60&70&80&90&100\\
         \hline
         \textbf{S} &0.78& 0.80&\textbf{0.84}& 0.73& 0.73& 0.71 & 0.72& 0.75\\
         \hline
    \end{tabular}
\end{table}

\begin{figure}
\vspace{-0.3cm}
\centering
  \includegraphics[width=0.9\columnwidth]{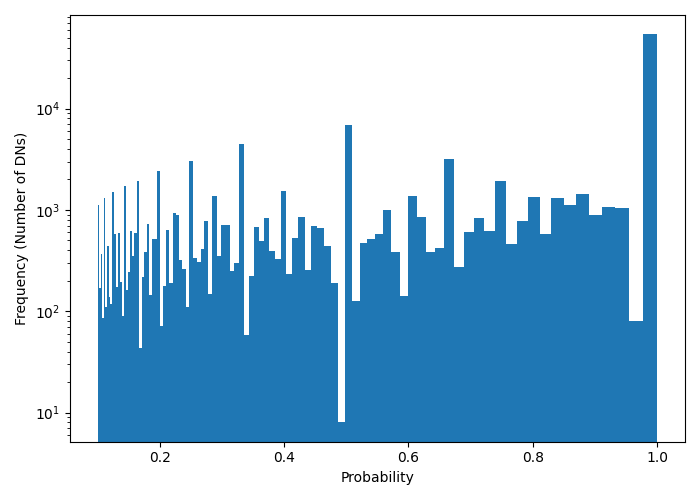}
\caption{Probability of being a malicious DN vs. Number of malicious DNs with that probability on a semi-log scale}
\label{figure12}
\vspace{-0.3cm}
\end{figure}

To capture the behavioral changes, we apply K-Means for different values of K$\in$[7,24] for each data segment present in the \emph{2H} granularity. Each data segment results in a different value of K for which silhouette score is maximum. From the clusters identified using the best K, for all the data segments, we select the cluster in which the number of malicious DNs are maximum. Within these selected clusters, we identify cosine similarity between the malicious and benign DNs and finally select those benign DNs where the cosine similarity $>$1-$\epsilon$ where $\epsilon$=10$^{-7}$. Using such a technique, for each DN, we get a series of labels representing, in a given granularity, how many times a particular DN was reported as malicious. Among all the DNs, we find 144930 DNs show malicious behavior at least once. Out of these DNs, 54114 DNs show persistent malicious behavior over time and have a high probability of being malicious. None of these DNs were marked malicious in any datasets we found. Figure~\ref{figure12} shows the histogram of the probability of previously unflagged DN being malicious. We then analyze how many identified DNs were marked and present in the new list of malicious DNs. This analysis reveals that none of the identified malicious DNs are present in the latest malicious blockchain DNs as of 7$^{th}$ April 2021. That is, the identified malicious DNs were not marked. This could be because of reasons such as \emph{(i)} the DN was not malicious or \emph{(ii)} it was able to evade detection by the authorities. Here, we do not reveal the identity of these DNs as we do not want to malign any dApp or BCW until authorities validate them. 




\vspace{-0.3cm}
\section{Conclusion}\label{sec:conc}

Safeguarding accounts in a permissionless blockchain is essential to minimize the risk of any fraudulent activity. Past approaches have identified that transactions performed by an account provide essential features using which we can detect malicious accounts. Besides the transactions, meta-information attached to the account also provides valuable inputs. In this work, we use Domain Name (DN) and related DNS query records to identify the malicious DNs and thus report malicious accounts associated with the particular DNs. Our approach captures the temporal aspects present in the DNS record to identify malicious DNs. In our work, besides the temporal aspects, we also use the non-temporal features of DNS query data. In the process, we also perform a comparative study of various techniques that identify malicious DNs.

Our results show that temporal features associated with a DN of a particular account can be useful towards the identification of whether that account is malicious or not. Our approach identifies 144930 DNs that show behavior as similar to known malicious DNs, and out of these 54114 DNs show persistent malicious behavior over time. None of these 54114 DNs were reported malicious in the newly identified malicious blockchain DNs as on 7$^{th}$ April 2021. This could be because the DN evaded the detection by the authorities, was not reported to the authorities or showed adversarial behavior. Integrating DN based features with other features (transaction and vulnerabilities-based features) would provide an edge in identifying malicious accounts in a blockchain. 

Further, continuous identification of malicious accounts and those accounts that show adversarial behavior based on such techniques is one possible extension. Deep learning algorithms can be used to detect malicious accounts based on DNs. Our technique uses association with underlying IP addresses. Identifying malicious DNs internally gives us IPs that are associated with malicious activities. Thus, resulting into blacklisting of associated IPs. Besides blockchain providers, security providing companies can also use such list to the safeguard underlying network. With respect to blockchain providers, nodes associated to such IPs and DNs can be blocked by ISPs. Nonetheless, we also warn users to not use predictable DNs. Such DNs reduce privacy and could reveal lot of information about the account~\cite{copeland20200}. With respect to ENS, one can use our approach on the Ethereum names to identify malicious accounts. However, in such a case DNS query-based, burst-based, and DNS graph-based features will not be applicable.

\vspace{-0.3cm}
\section*{Acknowledgement}\label{sec:ack}

This work is partially funded by the National Blockchain Project at IIT Kanpur sponsored by the National Cyber Security Coordinator's office of the Government of India and partially by the C3i Center funding from the Science and Engineering Research Board of the Government of India. We also thank Daniel Plohmann who made DGA dataset available.

\bibliographystyle{IEEEtran}
\bibliography{biblio.bib}

\end{document}